\documentclass[preprint]{revtex4}
\usepackage{graphicx}
\usepackage{amsmath}
\usepackage{hyperref}
\newcommand{\tpcut}{t^\prime_{cut}}
\newcommand{\tcut}{t_{cut}}
\newcommand{\etagk}{\eta_{\rm GK}}
\newcommand{\etace}{\eta_{\rm CE}}

\begin{document}

\title{The shear viscosity of quark-gluon matter calculated with parton transport and comparisons with the Chapman-Enskog results} 
\author{Mason Alexander Ross}
\author{Zi-Wei Lin}
\affiliation{Department of Physics, East Carolina University,
  C-209 Howell Science Complex, Greenville, NC 27858, USA}
\email[]{linz@ecu.edu}
\date{\today}

\begin{abstract}
We numerically calculate the shear viscosity of quark-gluon matter via the Green-Kubo relation with an improved ZPC model. We include all $2\leftrightarrow 2$ parton cross sections at finite temperature, which are  based on perturbative QCD and screened with thermal masses, and  consider massless  quark-gluon systems with Boltzmann statistics in chemical equilibrium. We then compare the Green-Kubo results with the analytical results from the leading-order Chapman-Enskog  method for the same parton cross sections over the temperature range $150-600$ MeV. We also examine the simpler case of isotropic and constant parton cross sections. Overall, we find that the two methods agree rather well. Specifically, the Green-Kubo results are greater than the Chapman-Enskog results by an average of $\sim 9\%$ for isotropic and constant cross sections and by an average of $\sim 3\%$ for finite-temperature pQCD cross sections, where the difference between the two methods is presumably due to higher-order corrections to the leading-order Chapman-Enskog results. 
\end{abstract}
\maketitle

\section{Introduction}

The shear viscosity of the quark-gluon plasma (QGP) is an important transport property that has attracted a lot of attention in recent years~\cite{Gyulassy:2004zy,Kovtun:2004de,Romatschke:2007mq}. 
In relativistic heavy ion collisions, the observed large anisotropic flows  indicate that the shear viscosity-to-entropy density ratio ($\eta/s$) of  the QGP is very small.
From the comparisons between the experimental data and 
theoretical models, especially hydrodynamics-based models, 
the QGP created in large collision systems at high energies 
is found to behave like a near perfect fluid~\cite{Schafer:2009dj,Song:2010mg,Bernhard:2019bmu,Parkkila:2021tqq}, 
and the extracted $\eta/s$ value near the QCD phase transition temperature $T_c$ is not too far from a conjectured lower limit of $1/(4\pi)$~\cite{Kovtun:2004de}. 

In certain cases, such as the early stage of heavy ion collisions at high energies, non-equilibrium dynamics may dominate before the system can be well described by hydrodynamics. There is also an open question on whether hydrodynamics is applicable for small collision systems or lower energies. To describe non-equilibrium dynamics, parton transport models have been developed and applied~\cite{Zhang:1997ej,Molnar:2000jh,Xu:2004mz,Lin:2004en,Cassing:2009vt,Kurkela:2022qhn}. 
Recently a QCD effective kinetic theory has also been developed~\cite{Kurkela:2018vqr}. Unlike hydrodynamics models, where the shear viscosity is an input, the shear viscosity of the QGP in transport models and kinetic theory is determined by the inter-species and intra-species interactions among partons.

Both analytical and numerical methods can be used to calculate the  shear viscosity of a quark-gluon matter. 
The Chapman-Enskog (CE) method for relativistic gases~\cite{Chapman1939TheMT,deGroot1980} 
provides a framework of calculating the shear viscosity of a single-species~\cite{Wiranata:2012br,Plumari:2012xz,MacKay:2022uxo} or multi-species~\cite{Wiranata:2013oaa} particle  system. 
Recently, the explicit first-order Chapman-Enskog expression for the shear viscosity of a massless quark-gluon matter, which is in chemical  equilibrium with Boltzmann statistics and subject to all and arbitrary $2\leftrightarrow 2$ parton cross sections, has been derived~\cite{Ohanaka:2026hjx}. 
This analytical expression involves only one- and two-dimensional integrals of the differential parton cross sections. 
Subsequently, this CE expression has been applied to given parton cross sections at finite temperature, which are based on perturbative QCD and screened with scaled thermal masses, to obtain the corresponding shear viscosity~\cite{Ohanaka:2026gla}. 
Numerically, one can take advantage of the Green-Kubo (GK) relation~\cite{Green1954JChPh..22..398G,Kubo:1957mj} and calculate the shear viscosity with a parton transport  model~\cite{Plumari:2012xz,Zhao:2020yvf,MacKay:2022uxo,Parisi:2025gwq}. The Green-Kubo relation calculates the shear viscosity of a system at or near thermal equilibrium as a numerical integral of the autocorrelation function of the shear components of the stress-energy tensor. 

In this study, we use the improved Zhang's parton cascade (ZPC)~\cite{Zhang:1997ej} and the Green-Kubo relation to calculate the shear viscosity of a massless quark-gluon matter in chemical equilibrium with Boltzmann statistics. 
The same parton cross sections at finite temperature 
as those used in the previous Chapman-Enskog  calculation~\cite{Ohanaka:2026gla} are used so that we can compare the shear viscosity results from the two methods.  
This paper is organized as follows. In Sec.~\ref{sec:2} we explain the method of numerically integrating the Green-Kubo autocorrelation function. In Sec.~\ref{sec:3} we calculate the shear viscosity of 
the quark-gluon matter under isotropic and energy-independent (i.e., constant) cross sections. In Sec.~\ref{sec:4} we calculate the shear viscosity of the quark-gluon matter subject to finite-temperature pQCD cross sections. After some discussions in  Sec.~\ref{sec:5}, we conclude in Sec.~\ref{sec:6}. 

\section{Methods}\label{sec:2}

We use the ZPC parton transport model~\cite{Zhang:1997ej}, which  numerically solves the relativistic Boltzmann equation
for $2\leftrightarrow 2$ scatterings with the cascade method and has been shown to accurately reproduce the time evolution of the single-particle distribution function
when using the $t$-minimum collision scheme~\cite{Zhao:2020yvf,Mendenhall:2025mwl}. 
For this work, we have extended ZPC to include
all $2\leftrightarrow 2$ scatterings among massless gluons and $N_f$ (anti)quark flavors. Additionally, we have incorporated into ZPC finite-temperature pQCD cross sections regulated by screening masses~\cite{Ohanaka:2026gla}. 

For a system at or near thermal equilibrium, the Green-Kubo formula for calculating the shear viscosity at a given temperature $T$ can be written as~\cite{Muronga:2003tb}
\begin{equation}
\eta = \frac{V}{T}\int_0^{\infty}\left<\bar{\pi}^{xy}(t+t^\prime) \bar{\pi}^{xy}(t^\prime)\right> dt,
\label{GK}
\end{equation}
where $V$ is the system volume, and $\left<.\right>$ represents the time ($t^\prime$) and ensemble average. 
The symbol $\bar{\pi}^{xy}(t)$ is the $xy$-component of the volume-averaged stress energy tensor of an event at time $t$, and for  discrete particles in ZPC it is calculated as~\cite{Zhao:2020yvf}
\begin{equation}
    \label{pi-xy}
    \bar{\pi}^{xy}(t)= \frac{1}{V}\sum_{i=1}^N \frac{p_i^xp_i^y}{E_i},
\end{equation}
where the sum is over all partons in the event. Note that the choice of $\pi^{xy}$ is not unique since any independent shear component of the tensor can be used in the Green-Kubo calculation for an isotropic parton matter in a box.  For the remainder of the paper, we adopt the  following notation for the autocorrelation function in Eq.\eqref{GK}:
\begin{equation}
C(t) \equiv \left<\bar{\pi}^{xy}(t+t^\prime)\bar{\pi}^{xy}(t^\prime)\right>.
\end{equation}
The time ($t^\prime$) and ensemble average of the correlation function is calculated in ZPC as~\cite{Zhao:2020yvf}
\begin{equation}
C(t)=\left<\frac{1}{\tpcut}\int_0^{\tpcut}\bar{\pi}^{xy}(t+t^\prime)\bar{\pi}^{xy}( t^\prime)dt^{\prime}\right>
\simeq\left<\frac{1}{N_t}\sum_{j=0}^{N_t-1}\bar{\pi}^{xy}(i\Delta t+j\Delta t)\bar{\pi}^{xy}(j\Delta t)\right>, 
\label{correlation-func}
\end{equation}
where $\tpcut=N_t\Delta  t$,  $N_t$ is the number of timesteps that we set to $400$ in this study, $t=i \Delta t$ and $t^\prime=j \Delta t$ after taking discrete timesteps, and the $\left < \cdot \right >$ symbols in Eq.\eqref{correlation-func}  represent the ensemble average.

Although Eq.\eqref{GK} involves an integral to infinite time, in practice one can only run the transport model simulation to finite time. As a result, we rewrite Eq.\eqref{GK} as
\begin{equation}
\eta = \frac{V}{T}
\left [ \int_0^{\tcut} C(t) dt+\int_{\tcut}^{\infty} C(t) dt\right ],
\label{GK2}
\end{equation}
where $\tcut$ is the cutoff time for the $C(t)$ calculation. 
The first part on the right-hand-side above can be calculated with numerical integration (e.g., using the trapezoidal rule), while the second  part represents the remainder that we can estimate. It has been observed that the autocorrelation function $C(t)$ decreases with  time roughly exponentially~\cite{Muronga:2003tb,Demir:2008tr,Wesp:2011yy,Plumari:2012xz,Zhao:2020yvf}. 
We find that for a single particle species the decrease of $C(t)$ with time is quite exponential~\cite{Zhao:2020yvf}; 
for a multi-species system, however, the decrease of $C(t)$ is less exponential at early times but more exponential at late times. 
Therefore, we assume that $C(t)$ beyond time $\tcut$ dampens exponentially as $e^{-t/\tau(\tcut)}$ and then estimate 
the integral in the remainder term of Eq.\eqref{GK2} as
\begin{equation}
\label{remainder}
\int_{\tcut}^{\infty} C(t) dt \simeq \tau(\tcut) \; C(\tcut).
\end{equation}
In the above, we extract the slope $\tau(\tcut)$ by fitting 
the two adjacent $C(t)$ values before (or at) the cutoff time $\tcut$ with an exponential function. 
In this study, we take the same value for $\tpcut$ and $\tcut$, thus  the transport model simulation is done over the time range $[0,2\,\tcut]$, and we determine their values so that the shear viscosity value from Eq.\eqref{GK2} changes little with further increase of $\tcut$. 
Note that prior to calculating the numerical integral, we scale the correlation function $C(t)$ of each event (with a constant very close to unity) so that the initial value $C(0)$ equals the theoretical expectation value 
\begin{equation}
\label{iniPiPi}
C(0) = \frac{4}{15}\frac{\epsilon T}{V},
\end{equation}
where $\epsilon=12(4+3N_f)T^4/\pi^2$ is the energy density of a system of massless gluons and $N_f$ (anti)quark flavors 
This scaling is done to reduce the event-by-event statistical fluctuation in the initial value of the correlation function.

\section{Shear Viscosity for Isotropic and Constant Cross Sections}\label{sec:3}

For a system of massless partons in chemical equilibrium interacting  with isotropic and energy-independent (i.e., constant) two-body cross sections,  
the first-order Chapman-Enskog expression for its shear viscosity is very explicit as it is given by~\cite{Ohanaka:2026hjx} 
\begin{equation}
\begin{aligned}
\etace^{\rm iso}&=\frac {12T}{5} \frac{A}{B},\\
A&=240 (2N_f \sigma_{gg} + \sigma_4) +8(112 + 48 N_f + 63 N_f^2) 
\sigma_{gq} +9 (48 + 32N_f + 27 N_f^2) \sigma_{q\bar q}^{gg}, \\ 
B&= 8 \left [8 \sigma_{gq}(28 \sigma_{gg} + 27N_f
\sigma_{gq})  + 3 \sigma_4 (20 \sigma_{gg} +  21 N_f \sigma_{gq}) \right ]  \\
&+ 9 (96 \sigma_{gg} + 176 N_f \sigma_{gq}+27 N_f
\sigma_4) \sigma_{q\bar q}^{gg} +   351 N_f (\sigma_{q\bar q}^{gg})^2. 
\label{etaIso}
\end{aligned}
\end{equation} 
In the above, we have written the elastic process $\sigma^{ij\rightarrow ij}$ as $\sigma_{ij}$ and the inelastic process $\sigma^{ij\rightarrow kl}$ as $\sigma_{ij}^{kl}$ for brevity. 
Note that Eq.\eqref{etaIso} involves seven independent cross sections, $\sigma_{gg}^{q\bar q}=9 \, \sigma_{q\bar q}^{gg}/32$ and thus it is not 
independent, and 
\begin{equation}
\sigma_4 \equiv \sigma_{qq} + \sigma_{q\bar q}+ (N_f-1)
(2\sigma_{qq^{\prime}} + \sigma_{q\bar q}^{q^{\prime} \bar q^{\prime}}).  
\label{sigma4}
\end{equation}

\begin{figure}[htb]
\includegraphics[width=0.6\linewidth]{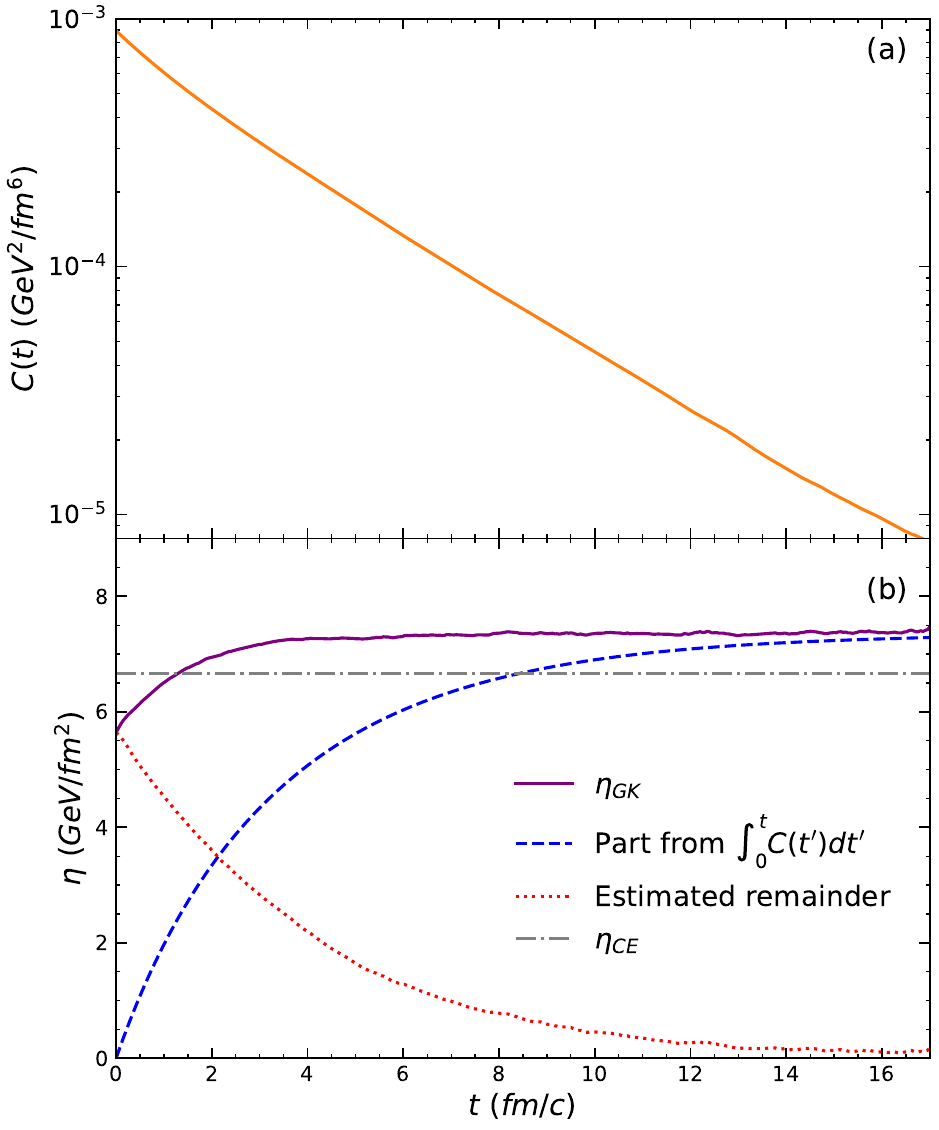}
\caption{(a) Autocorrelation function $C(t)$ versus time and (b) the shear viscosity versus cutoff time for an $N_f=1$ quark-gluon matter in chemical   equilibrium at $T=300$ MeV with isotropic cross sections $\sigma_{qq}=0.3$ fm$^2$  and  $\sigma_{q\bar{q}}^{gg}=0.2$ fm$^2$ as calculated with ZPC via the Green-Kubo relation.}
\label{ct}
\end{figure}

As an example, we show in Fig.~\ref{ct} the Green-Kubo results for the system of gluons and one flavor of (anti)quark, where panel (a) shows the autocorrelation  function $C(t)$ versus time $t$  and panel (b) shows the calculated shear viscosity as a function of the cutoff time. The system is set up in chemical and thermal equilibrium at  $T=300$ MeV, the partons are subject to isotropic scatterings with $\sigma_{qq}=0.3$ fm$^2$ and $\sigma_{q\bar{q}}^{gg}=0.2$ fm$^2$, and here we have simulated  90,000 events with 7000 partons  per event. In panel (a), an non-exponential form of the correlation  function is apparent at early times, thus a numerical integral is needed for an accurate determination of the shear viscosity, while  using the early-time slope of $C(t)$ only (as done in the case of single parton species~\cite{Zhao:2020yvf}) does not yield an accurate result.

In Fig.~\ref{ct}(b), we show the Green-Kubo shear viscosity (solid curve) as calculated from Eq.\eqref{GK2} as well as its two parts: the first part from the numerical integral (dashed) and the second part from the estimated remainder term (dotted). 
We see that the calculated shear viscosity $\etagk$ for a multi-species system depends on the cutoff time (when it is not long enough), and using too short a  cutoff time tends to underestimate the shear viscosity value. 
Empirically we observe that the calculated shear viscosity $\etagk$ becomes relatively stable after the autocorrelation function $C(t)$ has decreased by about an order of magnitude. 
On the other hand, the first part (from the integral of $C(t)$ from time zero to the cutoff time) becomes relatively stable much later, after the autocorrelation function $C(t)$ has decreased by about two orders of magnitude; this reflects the benefit of including the estimated remainder term in Eq.\eqref{GK2}. 
Naturally, as the cutoff time increases, the remainder term gradually approaches zero and the calculated $\etagk$ becomes rather stable. In this study, we take the average value of $\eta$ over the stable region as the shear viscosity value; for example, for the case shown in Fig.~\ref{ct}, the stable region is taken as $t \in [6,17]$ fm$/c$, resulting in $\etagk=7.34\pm0.06$ GeV/fm$^2$.
For comparison, the Chapman-Enskog shear viscosity $\etace^{\rm iso}=6.66$ GeV/fm$^2$ from Eq.\eqref{etaIso} is also shown (dot-dashed);  we see that the Green-Kubo value of the shear viscosity is $\sim  10\%$ greater than the CE value for this system.

\begin{figure}[htb]
\includegraphics[width=0.6\linewidth]{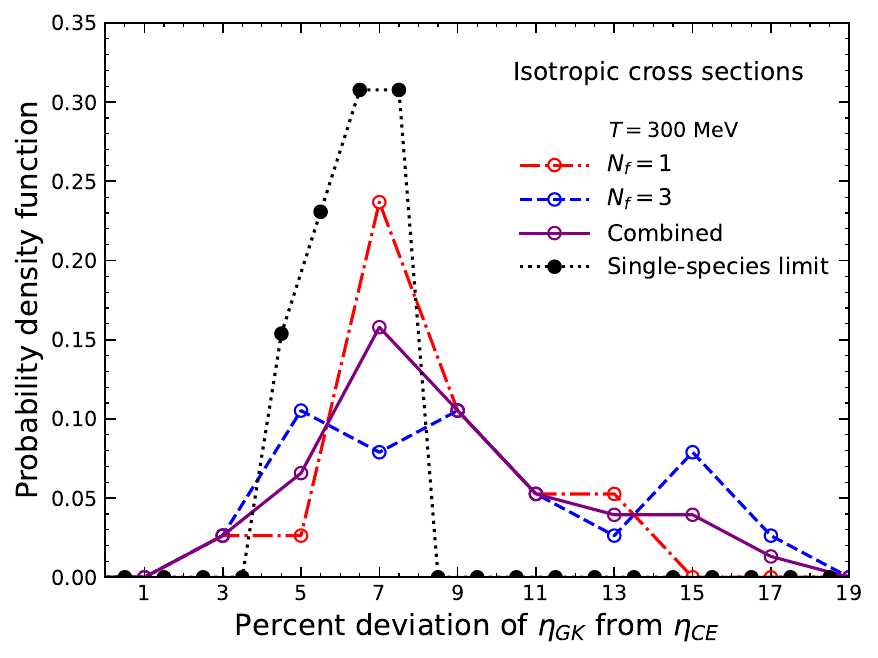}
\caption{The probability density functions of the percent deviation of $\etagk$ from $\etace$ for $N_f=1$ systems (dot-dashed), $N_f=3$ systems (dashed), and both of them (solid) at $T=300$ MeV under isotropic scatterings  with different sets of $2\leftrightarrow 2$ cross sections; the black dotted curve represents the PDF for the configurations in Table~\ref{ssl} used to check the single-species limit.}
\label{iso-pdf-withSSL}
\end{figure}

To better assess the difference between the GK and CE results on the shear viscosity, we have performed the calculations for many different systems. Specifically, we have considered twenty-five $N_f=1$ systems and twenty $N_f=3$ systems with various isotropic cross sections within the following ranges:  $\sigma_{gg},\sigma_{gq},\sigma_{qq}, \sigma_{q\bar {q}} ~\&~ \sigma_{q\bar{q}}^{gg} \in [0,0.3]$ fm$^2$, $\sigma_{qq^\prime}\in[0,0.1]$ fm$^2$, and $\sigma_{q\bar{q}}^{q^\prime\bar{q}^\prime} \in[0,0.2]$ fm$^2$. 
We show in Fig.~\ref{iso-pdf-withSSL} the probability density functions (PDFs) of the percent deviation of the $\etagk$ value from the $\etace$ value for these configurations. We find that the average deviations for these $N_f=1$ and $N_f=3$ systems are $8.2\%$ and $9.5\%$, respectively, while the average deviation for both systems is $8.8\%$, meaning that the Green-Kubo value for the shear viscosity is typically $\sim 9\%$ greater than the leading-order Chapman-Enskog value. 
We also see that the spread of the deviation values is rather big with a standard deviation $\sim 3\%$. 
The difference between the $\etagk$ and $\etace$ values is presumably due to higher-order corrections to the leading-order Chapman-Enskog result, since it is already known that higher-order terms increase the leading-order $\etace$ value by $5.7\%$ for a single particle species under isotropic scatterings~\cite{Wiranata:2012br}.
Although higher-order Chapman-Enskog corrections have not been derived for multi-species systems, the deviations between the $\etagk$ values and the leading-order $\etace$ values observed in this study provide numerical  information. For example, we find that all systems shown in Fig.~\ref{iso-pdf-withSSL} with deviation greater than $12\%$ have a collision rate per gluon that is at least a factor of five different from the collision rate per quark.

Since the effect of higher-order Chapman-Enskog terms is known for a single particle species under isotropic scatterings~\cite{Wiranata:2012br}, we perform the Green-Kubo calculation for a single particle species (gluon gas) at $T=300$ MeV with an isotropic elastic cross section $\sigma=0.04$ fm$^2$. We have also calculated ten different multi-species configurations as listed in  Table~\ref{ssl}, each of which should have the same shear viscosity as the above single species case is based on the general single-species limit for the shear viscosity proposed earlier~\cite{Ohanaka:2026hjx}.
Since the time evolution of the autocorrelation function in Eq.\eqref{GK}  solely depends on the parton momentum transfer, a multi-species system where every particle species has the same collision rate per particle will have the same shear viscosity of a single-species system at the corresponding effective cross section~\cite{Ohanaka:2026hjx}:
\begin{equation} 
\sigma_{\rm eff}=\frac{4\sigma_{gg} + 3N_f \sigma_{gq} + 4 N_f
  \sigma_{gg}^{q\bar  q} }{4 + 3 N_f},
\label{sigmaIso}
\end{equation}
provided that the scatterings have the same angular distribution. 
Under isotropic and constant $2\leftrightarrow 2$ cross sections, the collision rate per parton for gluons and each (anti)quark species for a quark-gluon system in chemical equilibrium are given respectively by~\cite{Ohanaka:2026hjx}
\begin{equation}
r_g/n = x_g\sigma_{gg}+2N_fx_q\sigma_{gq}+N_fx_g\sigma_{gg}^{q\bar{q}},\  r_q/n = r_{\bar{q}}/n=x_g\sigma_{gq}+x_q\sigma_4+x_q\sigma_{q\bar{q}}^{gg}, 
\label{rates}
\end{equation}
where $n$ is the total number density of all partons, 
the number fractions of gluons and each quark or antiquark species are respectively given by 
\begin{equation}
x_g=\frac{4}{4 + 3N_f}, ~x_q= \frac{3\;(1-\delta_{0N_f})}{8 + 6N_f}, 
\label{xi}
\end{equation}
and setting $r_q=r_g$ reduces the system to the single-species case. 
As shown in Table~\ref{ssl}, we have considered five configurations of $N_f=1$ systems and five configurations for $N_f=3$ systems at $T=300$ MeV, 
each of which is expected to have the same shear viscosity as a single particle species with the cross section of $0.04$ fm$^2$ (at the same temperature). The $\sigma_4$ term defined in Eq.\eqref{sigma4} contains four separate cross sections. Specifically, we set  all four cross sections in $\sigma_4$ to zero for the $N_f=1$ calculations in  Table~\ref{ssl}. For the $N_f=3$ calculations, we set $\sigma_4=\sigma_{qq}$  (with the other cross sections in $\sigma_4$ set to zero) for run 6, 
we set $\sigma_4=\sigma_{q\bar{q}}$ for run 7, 
we set $\sigma_4=\sigma_{q\bar{q}}/2  + \sigma_{qq}/2$ with $\sigma_{q\bar{q}}=\sigma_{qq}$ for run 8, 
we set $\sigma_4=\sigma_{q\bar{q}}  + 4\sigma_{qq^\prime}$ with  $\sigma_{q\bar{q}}=\sigma_{qq^\prime}$ for run 9, and we set $\sigma_4=\sigma_{qq}  + 2\sigma_{q\bar{q}}^{q^\prime\bar{q}^\prime}$ with $\sigma_{qq}=\sigma_{q\bar{q}}^{q^\prime\bar{q}^\prime}$ for run 10.

\begin{table}
\begin{center}
\begin{tabular}{|c|c|c|c|c||c|c|c|c|c|}
\hline
\multicolumn{5}{|c||}{$N_f=1$} & \multicolumn{5}{c|}{$N_f=3$}\\
\hline
Run $\#$ &$\sigma_{gg}$ & $\sigma_{gq}$ &$\sigma_{q\bar{q}}^{gg}$ & $\sigma_{4}$ &Run $\#$ &$\sigma_{gg}$ & $\sigma_{gq}$ &$\sigma_{q\bar{q}}^{gg}$ & $\sigma_{4}$\\
\hline \hline
1 & 0.0175 & 0.0700 &0.0000  & 0.0000 & 6 & 0.0250 & 0.0467 &0.0000 & 0.2222 \\
2 & 0.0175 & 0.0525 & 0.0467 & 0.0000 &7 & 0.0200 & 0.0414 & 0.0200 & 0.2163 \\
3 & 0.0175 & 0.0350 & 0.0933 & 0.0000 &8 & 0.0100 & 0.0496  & 0.0100 & 0.2044 \\
4 & 0.0175 & 0.0175 & 0.1400 & 0.0000 &9 & 0.0050 & 0.0331 & 0.0600 & 0.1985 \\
5 & 0.0175 & 0.0000 & 0.1867 & 0.0000 & 10 & 0.0000 & 0.0203 & 0.1000 & 0.1926 \\
\hline
\end{tabular}
\end{center}
\caption{Configurations for $N_f=1$ and $N_f=3$ quark-gluon systems to test the single-species limit; all cross sections are isotropic in the unit of fm$^2$, and all configurations should have the same shear viscosity as a single particle species with $\sigma_{\rm eff}=0.04$ fm$^2$ at the same temperature.}
\label{ssl}
\end{table}

\begin{figure}[htb]
\includegraphics[width=0.6\linewidth]{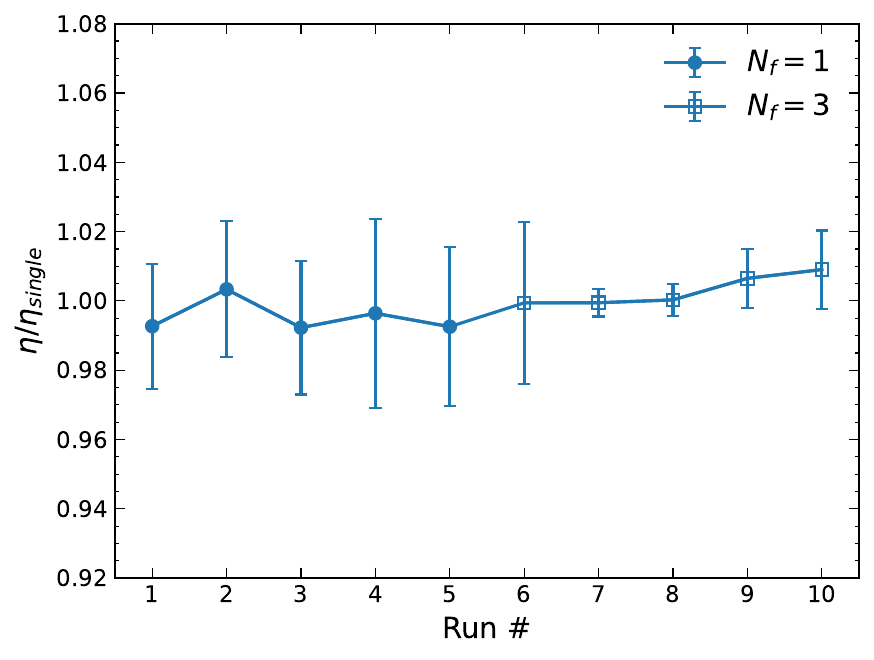}
\caption{Ratio of the shear viscosity of various 
single-species-limit configurations in Table~\ref{ssl}  over that of a gluon gas at $\sigma=0.04$ fm$^2$ at the same temperature $T=300$ MeV.}
\label{SSL-ratio}
\end{figure}

Figure~\ref{SSL-ratio} shows the ratio of the Green-Kubo shear viscosity value from each configuration  in Table~\ref{ssl} over the value for the true single-species case at $\sigma=0.04$ fm$^2$ (denoted as $\eta_{single}$). We see that indeed each configuration has same shear viscosity as the single species case within error, regardless of the number of species present or the value of  individual cross sections. 
This also indirectly validates as our extension of the ZPC model to all $2\leftrightarrow 2$ parton scatterings. 
In addition, the shear viscosity value for the gluon gas at $T=300$ MeV and $\sigma=0.04$ fm$^2$ is $\eta_{single}=9.6 \pm 0.2$ GeV/fm$^2$, which is $7\pm2\%$ bigger than the leading-order CE result $\etace=6T/5\sigma=9.0$ GeV/fm$^2$; 
this difference is consistent with the previously-known effect of higher-order CE corrections for a single particle species. 
The probability density function of the percent deviation of $\etagk$  from the $\etace$ value for these single-species limit configurations  
is shown in Fig.~\ref{iso-pdf-withSSL} (black dotted curve),  
which is much narrower than the distributions for the other isotropic configurations. The mean deviation from $\etace$ is $6.3\%$; 
this is rather consistent with the previous finding~\cite{Wiranata:2012br} that higher-order terms (up to the $16^{th}$ order) increase the  leading-order $\etace$ value by $5.7\%$ for a single particle species under isotropic scatterings. 

\section{Shear Viscosity for Finite-Temperature pQCD Cross Sections}\label{sec:4}

In the extended ZPC model, we have implemented leading-order pQCD finite-temperature cross sections. 
Unlike the AMY framework~\cite{Arnold:2000dr,Arnold:2003zc} that screens the two-body matrix elements with parton thermal self-energies, we take the simpler approach by screening with scaled Debye mass and thermal Fermion mass~\cite{Ohanaka:2026gla}:
\begin{equation}
\mu_D =\xi m_D= \xi \sqrt{1+\frac{N_f}{6}}\;gT, 
~\mu_F = \xi m_F= \xi \frac{gT}{\sqrt{6}},
\end{equation}
where $\xi$ is the scaling factor in the ``$\xi$-prescription'' of screening~\cite{AbraaoYork:2014hbk,Boguslavski:2024jwr}. 
We take the following leading-order strong coupling~\cite{ParticleDataGroup:2014cgo}
\begin{equation}
\label{coupling}
\alpha_s(Q^2) = \frac{4\pi}{\left(11-\frac{2}{3}N_f\right) \ln \left(Q^2/\Lambda^2\right)},
\end{equation}
where $g^2=4\pi \alpha_s$. In this study, we choose $\Lambda=250$ MeV~\cite{Ohanaka:2026gla}, we take $Q=3T$~\cite{Ghiglieri:2018dib} 
or $Q=2\pi T$~\cite{Blaizot:2001tn} to study the effect of the renormalization scale $Q$ on the shear viscosity, 
and we take the scaling factor $\xi=1$ by default unless specified otherwise. 
It has been shown that $\xi=\sqrt{0.4}$ allows the leading-order Chapman-Enskog shear viscosity to be very close to the AMY result for two-body scatterings at finite temperature for Boltzmann statistics~\cite{Ohanaka:2026gla}. Therefore, we have also made calculations at $\xi=\sqrt{0.4}$. 

We now consider a massless quark-gluon matter in chemical equilibrium at $N_f=3$ in this Section. We take the same differential cross sections for $2\leftrightarrow 2$ parton scatterings as in a previous Chapman-Enskog calculation~\cite{Ohanaka:2026gla}, which are given as follows.

$(gg\rightarrow gg)$:
\begin{equation}
    \frac{d\sigma}{dt} = \frac{9g^4}{32\pi s^2} \left[\frac{3s}{s+\mu_D^2}-\frac{su}{(t-\mu_D^2)^2}-\frac{st}{(u-\mu_D^2)^2}-\frac{tu}{s(s+\mu_D^2)}\right].
\label{dsdt1}
\end{equation}

$(gq\rightarrow gq)$ and ($g\bar{q}\rightarrow g\bar{q})$:
\begin{equation}
    \frac{d\sigma}{dt} = \frac{g^4}{48\pi s^2}\left[3\frac{s^2+u^2}{(t-\mu_D^2)^2} - \frac{4}{3}\left(\frac{s}{u-\mu_F^2} + \frac{u}{s+\mu_F^2}\right)\right].
\end{equation}

$(qq\rightarrow qq)$ and $(\bar{q}\bar{q}\rightarrow \bar{q}\bar{q})$:
\begin{equation}
    \frac{d\sigma}{dt} = \frac{g^4}{36\pi s^2}\left[\frac{s^2+u^2}{(t-\mu_D^2)^2} +\frac{s^2+t^2}{(u-\mu_D^2)^2} -\frac{2}{3}\frac{s^2}{(t-\mu_D^2)(u-\mu_D^2)} \right].
\end{equation}

$(q\bar{q}\rightarrow q\bar{q})$:
\begin{equation}
    \frac{d\sigma}{dt} = \frac{g^4}{36\pi s^2}\left[\frac{s^2+u^2}{(t-\mu_D^2)^2} + \frac{t^2+u^2}{s(s+\mu_D^2)}-\frac{2}{3}\frac{u^2}{s(t-\mu_D^2)}\right].
\end{equation}

$(qq^\prime\rightarrow qq^\prime), $ $(q\bar{q}^\prime\rightarrow q\bar{q}^\prime), $ $(q^\prime\bar{q}\rightarrow q^\prime\bar{q})$, and  $(\bar{q}\bar{q}^\prime\rightarrow \bar{q}\bar{q}^\prime)$:
\begin{equation}
    \frac{d\sigma}{dt} = \frac{g^4}{36\pi s^2}\frac{s^2+u^2}{(t-\mu_D^2)^2}.
\end{equation}

$(q\bar{q}\rightarrow q^\prime\bar{q}^\prime)$:
\begin{equation}
\frac{d\sigma}{dt} = \frac{g^4}{36\pi s^3} \frac{t^2+u^2}{s+\mu_D^2}.
\end{equation}

$(gg\rightarrow q\bar{q})$:
\begin{equation}
    \frac{d\sigma}{dt} = \frac{g^4}{128\pi s^2}\left[\frac{4}{3}\left(\frac{t}{u-\mu_F^2}+\frac{u}{t-\mu_F^2}\right)-\frac{3(t^2+u^2)}{s(s+\mu_D^2)}\right].
\end{equation}

$(q\bar{q}\rightarrow gg)$:
\begin{equation}
    \frac{d\sigma}{dt} = \frac{g^4}{18\pi s^2}\left[\frac{4}{3}\left(\frac{t}{u-\mu_F^2}+\frac{u}{t-\mu_F^2}\right)-\frac{3(t^2+u^2)}{s(s+\mu_D^2)}\right].
\label{dsdt2}
\end{equation}
In the above, $s,t,u$ are the standard Mandelstam variables, and $q$ and $q^\prime$ represent quarks of different flavors. 
Note that the cross section for process $ij \rightarrow kl$ is given by~\cite{Ohanaka:2026hjx} 
\begin{equation}
\sigma^{ij \rightarrow kl}=\frac{1}{1+\delta_{kl}} \int_{-s}^0 {\frac {d\sigma}{dt}}^{ij \rightarrow kl} dt.
\end{equation}

\begin{figure}[htb]
\includegraphics[width=0.6\linewidth]{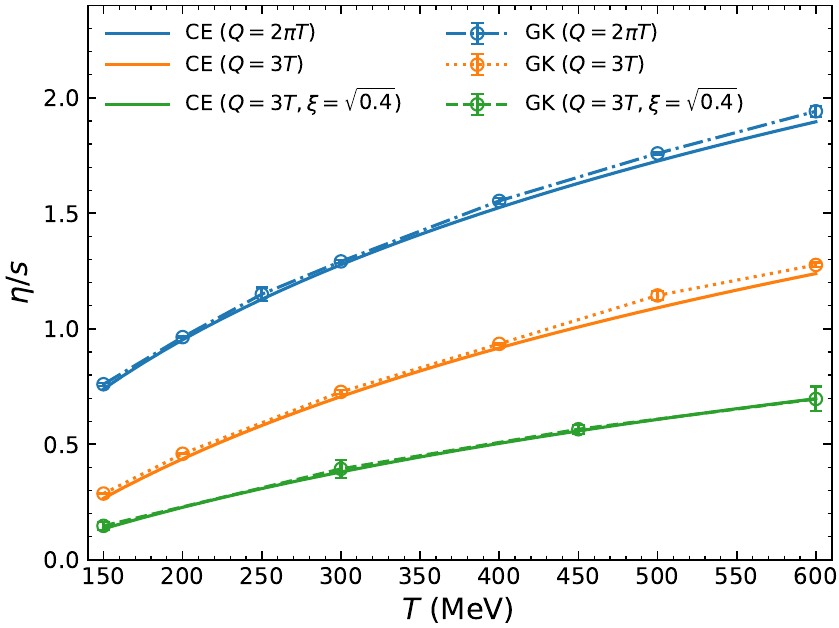}
\caption{The $\eta/s$ ratio of a $N_f=3$ quark-gluon matter under finite-temperature pQCD  cross sections versus temperature; both Green-Kubo and leading-order Chapman-Enskog results are shown for three parameters sets: 
$Q=2\pi T ~\&~ \xi=1$, $Q=3T ~\&~ \xi=1$, and $Q=3T ~\&~ \xi=\sqrt{0.4}$.}
\label{etas-vs-T}
\end{figure}

Figure~\ref{etas-vs-T} shows the shear viscosity-to-entropy density ratios in the temperature range 150-600 MeV from the numerical Green-Kubo method (curves with symbols) and the analytical leading-order Chapman-Enskog method (curves without symbols). 
Note that the entropy density $s$ for Boltzmann statistics is given by $s=4n=16(4+3N_f)T^3/\pi^2$, 
and the explicit analytical leading-order Chapman-Enskog expression of shear viscosity for such a quark-gluon system has been derived recently~\cite{Ohanaka:2026hjx} and then applied to 
finite-temperature pQCD  cross sections given by Eqs.\eqref{dsdt1}-\eqref{dsdt2}~\cite{Ohanaka:2026gla}. 
The general Chapman-Enskog expression is given by
\begin{equation}
\etace = 160\, T ~ \frac{x_g^2 \left[C_{qq}^{00} + C_{q\bar{q}}^{00} +
    2(N_f-1)C_{qq^{\prime}}^{00}\right] - 4 N_f x_g x_q\, C_{gq}^{00} +2 N_f
  x_q^2 \,C_{gg}^{00}} {C_{gg}^{00}\left[C_{qq}^{00} +
    C_{q\bar{q}}^{00} + 2(N_f-1)C_{qq^{\prime}}^{00}\right] - 2N_f
  \left(C_{gq}^{00}\right)^2},
\label{etaNf}
\end{equation}
where each matrix elements $C_{ij}^{00}$ involves only one- and two-dimensional integrals of the differential parton cross sections~\cite{Ohanaka:2026hjx}. 
We calculate three different configurations with different values for the renormalization scale $Q$~\cite{Ghiglieri:2018dib} and scaling factor $\xi$~\cite{Ohanaka:2026gla}:
$Q=2\pi T ~\&~ \xi=1$ (dot-dashed), $Q=3T ~\&~ \xi=1$ (dotted), and $Q=3T ~\&~ \xi=\sqrt{0.4}$ (dashed). We see that the $\eta/s$ ratios versus temperature from the Green-Kubo method are very close to  those from the leading-order Chapman-Enskog method (solid curves) for all three configurations. The GK result is slightly bigger than the leading-order CE result, consistent with the expectation that higher-order CE terms would slightly increase the CE shear viscosity. 
In addition, the $\eta/s$ ratio is seen to depend strongly on the choice of $Q$ and $\xi$. Decreasing $Q$ from $2\pi T$ to $3T$ increases the strong coupling, the parton cross sections, and consequently decreases the shear viscosity; while decreasing the $\xi$ value has a similar effect. 
As a result, the configuration with $Q=3T ~\&~ \xi=\sqrt{0.4}$ 
gives the lowest $\eta/s$ values in Fig~\ref{etas-vs-T}, 
with $\etagk/s \sim 0.15$ near the QCD phase transition temperature $T_c \simeq 156$ MeV~\cite{HotQCD:2018pds}. 

\begin{figure}[htb]
\includegraphics[width=0.6\linewidth]{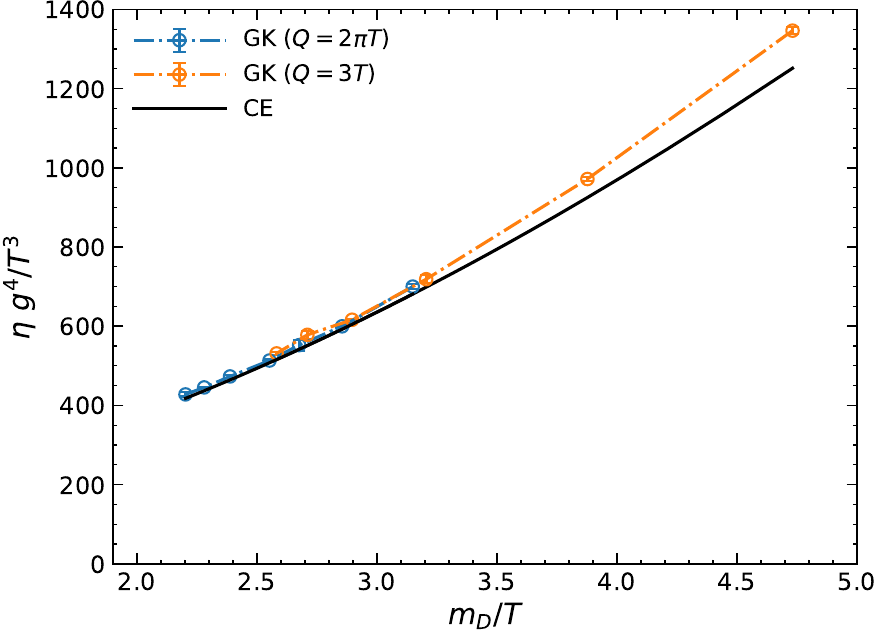}
\caption{Shear viscosity scaled by $g^4/T^3$ as a function of $m_D/T$ for 
$N_f=3$ quark-gluon gas under finite-temperature pQCD  cross sections with 
$Q=2\pi T$ and $Q=3T$ from the Green-Kubo relation in comparison with the leading-order Chapman-Enskog results.}
\label{eta-vs-md}
\end{figure}

Figure~\ref{eta-vs-md} presents the shear viscosity results by plotting the $\eta g^4/T^3$ values as functions of $m_D/T$; this way the result does not depend on the renormalization scale $Q$~\cite{Arnold:2000dr,Arnold:2003zc}.  
We show the Green-Kubo results for both $Q=2\pi T$ and $Q=3T$ (with $\xi=1$), and indeed the two results form a continuous curve within error. 
Note that this $Q$-independence has been shown with the Chapman-Enskog results on the shear viscosity~\cite{Ohanaka:2026gla}. 
In addition, the Green-Kubo results in Fig.~\ref{eta-vs-md} are several percent greater than the leading-order CE results as expected. 
Note that for the same temperature, different $Q$ values 
correspond to different $m_D/T$ values; for example, $m_D/T$ is 3.15 for $Q=2\pi T$ but is 4.73 for $Q=3T$ at $T=150$ MeV.  

\begin{figure}[htb]
\includegraphics[width=0.6\linewidth]{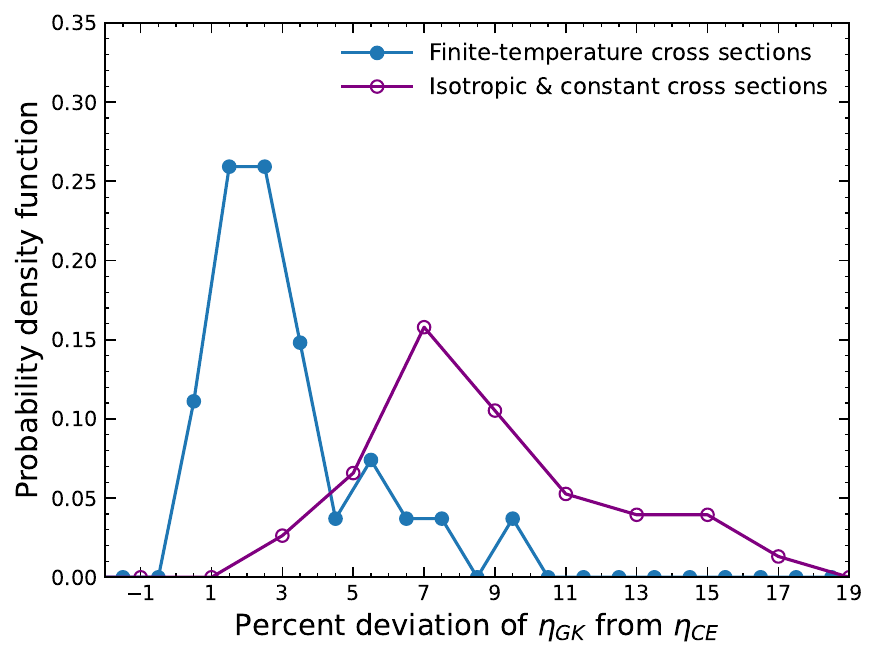}
\caption{Probability density functions of the percent deviation of the $\etagk$ value from the leading-order $\etace$ value for finite-temperature cross sections and for isotropic \& constant cross sections.}
\label{pdf}
\end{figure}

We show in Fig.~\ref{pdf} the probability density functions of the percent deviation of the $\etagk$ value from the leading-order $\etace$ value for finite-temperature cross sections and for isotropic \& constant cross sections. We see that the $\etagk$ value is no lower than the  leading-order $\etace$ value for all the cases in the figure, 
where $\etagk$ is on average closer to the leading-order $\etace$ value for finite-temperature cross sections (mean deviation at $\sim 3\%$  with individual deviations ranging from 0 to 10\%) than for isotropic \& constant cross sections (mean deviation at $\sim 9\%$ with individual deviations ranging  from $\sim 4\%$ to 17\%). This indicates that the higher-order CE corrections are smaller for  finite-temperature cross sections than for isotropic \& constant cross sections. 

\begin{figure}[htb]
\includegraphics[width=0.6\linewidth]{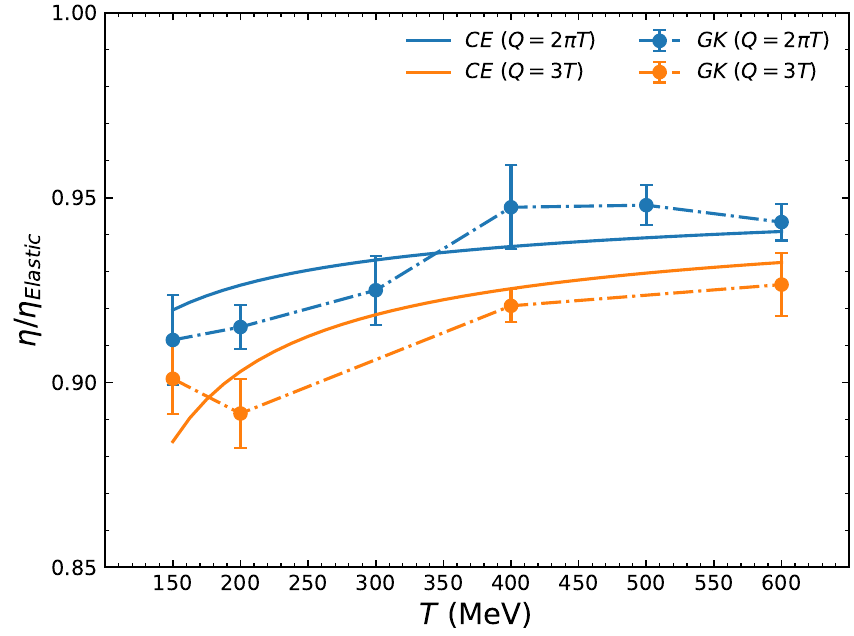}
\caption{Ratio of the shear viscosity to the shear viscosity from
only elastic scatterings for $N_f=3$ quark-gluon matter under finite-temperature cross sections from the Green-Kubo method and the leading-order Chapman-Enskog method for $Q=2\pi T$ and $Q=3T$ (with $\xi=1$).}
\label{ratio-elastic}
\end{figure}

The effects of inelastic $2\leftrightarrow 2$ collisions 
are shown in Fig.~\ref{ratio-elastic}, which plots the ratio of the full shear viscosity (i.e., from both elastic and inelastic collisions) to the shear viscosity from elastic collisions only. Both the Green-Kubo results from ZPC  calculations and the  leading-order  CE results are shown. 
We see that inelastic $2\leftrightarrow 2$ collisions, including $gg \leftrightarrow q\bar q$ and  $q\bar q \leftrightarrow q^\prime \bar {q^\prime}$, reduce the shear viscosity by $\sim 5-11\%$, with a larger effect at lower temperatures. 
The GK results are seen to be mostly consistent with the CE results. In addition, we see that the inelastic effect depends on the $Q$  value, while the effect has been shown with the Chapman-Enskog method to also depend on the values of $N_f$ and $\xi$~\cite{Ohanaka:2026gla}. 

\section{Discussions}\label{sec:5}

\begin{figure}[htb]
\includegraphics[width=0.8\linewidth]{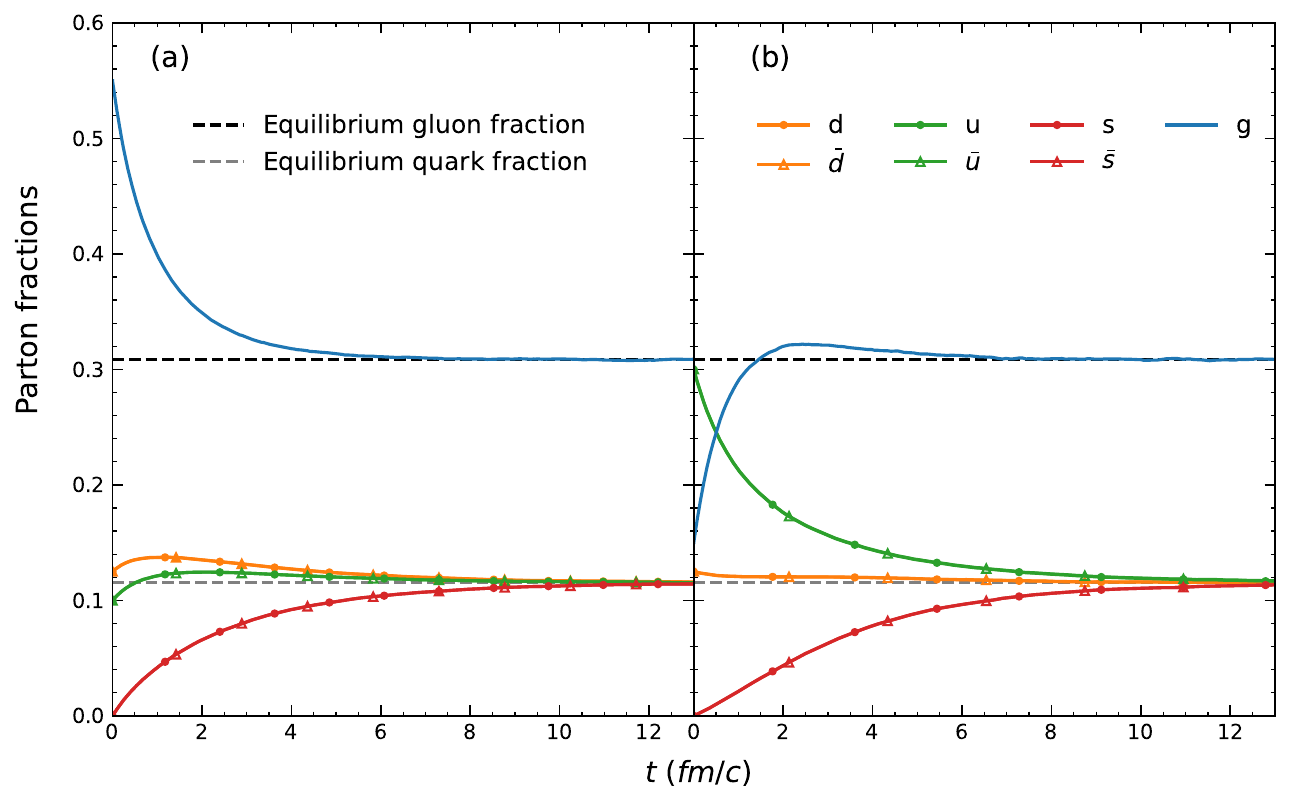}
\caption{Time evolutions of parton fractions of two $N_f=3$  quark-gluon systems with different initial fractions with both systems under finite-temperature cross sections at $T=150$ MeV, $Q=3T$, and $\xi=1$; dashed lines indicate the expected fractions in chemical equilibrium.
}
\label{equilibration}
\end{figure}

We have extended the ZPC model~\cite{Zhang:1997ej} to include all $2\leftrightarrow 2$ parton scatterings for this study. 
To further verify the improved model, we can check the chemical equilibration by calculating the relative populations of partons in ZPC as a function of time starting from chemical non-equilibrium. 
Figure~\ref{equilibration} shows our test results for $N_f=3$ quark-gluon matter at $T=150$ MeV with finite-temperature cross sections at  $Q=3T$ and $\xi=1$, where at time zero the system in panel (a) has overpopulated gluons and the system in panel (b) has underpopulated gluons. 
For both cases, the chemical equilibration is quite fast in the first couple of fm$/c$ and then the fractions gradually approach the equilibrium values. 
We also see that in panel (a) the fraction of $u$ (and $\bar{u}$) quarks starts below its chemical equilibrium value and then exceeds the equilibrium value before saturating at the equilibrium value; this interesting feature is also seen for gluons in panel (b). In both cases, the system almost reaches complete chemical equilibrium after $\sim 10$ fm$/c$. 
Note that the inelastic cross sections and thus the chemical equilibration time strongly depend on the temperature and 
the values of $Q$ and $\xi$, 
where a smaller value of $Q$ (and $\xi$) will lead to faster chemical equilibration~\cite{Ohanaka:2026gla}. 

We have seen that the autocorrelation function $C(t)$ in Fig.~\ref{ct} 
at early times is concave upward in the log-linear plot; empirically we find that this is the case for most configurations that we have computed with the ZPC model. This feature would be natural if the correlation function can be considered as a sum of multiple (positive) exponetially-decreasing functions. 
Indeed, for a multi-species system, particles of different species often have different scattering rates and thus different relaxation times; for example, for the configuration shown in Fig.~\ref{ct}, the collision rate per parton for gluons is roughly five times that for (anti)quarks. More studies will be needed to better understand the composition and time-dependence of the autocorrelation function, which may be necessary for  extending the Green-Kubo method to more cases such as non-chemical equilibrium.

For a wide range of cross section configurations for the quark-gluon matter, Fig.~\ref{pdf} shows that the Green-Kubo shear viscosity is greater than the leading-order Chapman-Enskog value by  $2-18$\% for the case of isotropic \& constant cross sections or by $0-10$\% for the case of finite-temperature cross sections. This suggests that the effect of higher-order Chapman-Enskog  corrections varies. Our preliminary results suggest that the effect depends on the collision rate difference among different particle species, but further studies will be needed to better understand the magnitudes of higher-order Chapman-Enskog corrections for a multi-species system.

\section{Conclusion}\label{sec:6}

We use an improved ZPC parton transport and the Green-Kubo relation to numerically calculate the shear viscosity of massless quark-gluon matter in chemical equilibrium with Boltzmann statistics under all $2\leftrightarrow 2$ scatterings. We then compare the Green-Kubo results with the leading-order Chapman-Enskog  results for the same parton cross sections. 
We include all $2\leftrightarrow 2$ parton cross sections, including the simpler case of isotropic \& constant parton cross sections and the more realistic finite-temperature cross sections that are based on perturbative QCD and screened with scaled thermal masses. Overall, we find that the two methods agree rather well. Specifically, the Green-Kubo values of shear viscosity are  greater than the Chapman-Enskog values by an average of $\sim 9\%$ for isotropic \& constant cross sections and by an average of $\sim 3\%$ for  finite-temperature pQCD cross sections over the temperature range of $150-600$ MeV. The difference between the two methods is presumably due to higher-order Chapman-Enskog corrections, which are known to increase the leading-order Chapman-Enskog  value by $5.7\%$ for a single particle species under isotropic scatterings. These results provide confirmation of the analytical leading-order Chapman-Enskog expression for the shear viscosity of massless quark-gluon matter. They also help validate the improved ZPC model, which has been extended to include all $2\leftrightarrow 2$ parton scatterings with finite-temperature pQCD cross sections. This study lays the foundation for shear viscosity calculations in more general cases such as particles with finite masses and non-chemical equilibrium.

\section*{Acknowledgments}
We thank G. Moore for helpful discussions on screening prescriptions. This work has been supported by the National Science Foundation under Grant No. 2310021.

\bibliography{refs}

\end{document}